\documentclass[12pt,epsf,amstex]{article}
\usepackage [dvips]{graphicx}
\usepackage{amsmath}
\usepackage{times}
\usepackage{psfig}
\usepackage{epsfig}
\usepackage{amssymb}

\newcommand{\dd}{\text{d}}

\newcommand{\ee}{\text{e}}

\newcommand{\p}{\partial}

\newcommand{\eps}{\varepsilon}

\newcommand{\bp}{\text{\bf p}}

\begin{document}
\begin{center}
{\Large{\bf Large deviations in weakly interacting\\
\vspace{8pt}
boundary driven lattice gases}}
\end{center}

\vspace{2cm}
\begin{center}
{Fr\'ed\'eric van Wijland$^{1,2}$ and Zolt\'an R\'acz$^{1,3}$}
\end{center}

\noindent ${}^1$Laboratoire de Physique Th\'eorique, B\^at. 210,
Universit\'e de Paris-Sud,\\ 91405 Orsay Cedex, France.\\

\noindent ${}^2${P\^ole Mati\`ere et Syst\`emes Complexes, Universit\'e de Paris VII, France.\\

\noindent ${}^3$Institute for Theoretical Physics, HAS Research Group,
E\"otv\"os University \\1117 Budapest, P\'azm\'any s\'et\'any 1/a, Hungary.\\
}\\\\

\begin{center}{\bf Abstract}\\
\end{center}
One-dimensional, boundary-driven lattice gases with local interactions
are studied in the weakly interacting limit. The density profiles
and the correlation functions are calculated to first order in the interaction
strength for zero-range and short-range processes differing only
in the specifics of the detailed-balance dynamics. Furthermore,
the effective free-energy (large-deviation function) and the integrated
current distribution are also found to this order. From the former, we find
that the boundary drive generates long-range correlations only for
the short-range dynamics while the latter provides support to an
additivity principle recently proposed by Bodineau and Derrida.

\vskip 2cm
\newpage

\section{Introduction}

One-dimensional boundary driven lattice gases are simple model systems which
allow detailed studies of nonequilibrium steady-states.
They are relevant in the sense that steady states in experiments
are often produced by choosing appropriate boundary conditions
(heating a horizontal layer of liquid from below being the most widely quoted
example) and, furthermore, these models appear to capture some
important consequences of the nonequilibrium drive such as e.g. the
generation of long-range correlations.

Analytical approaches to driven lattice gases proceed along two distinct paths.
On one hand, exploiting the matrix product method, Derrida and
coworkers \cite{{derridalebowitzspeer1},{derridalebowitz},{derridalebowitzspeer2},
{derridalebowitzspeer3},{derridaenaud},{derridadoucotroche}} have come up with a
series of exact results for the exclusion process
(hard-core particles undergoing symmetric or
biased diffusive motion) in one dimension .
On the other hand, in a parallel series of papers,
Bertini and coworkers
\cite{{bertinisolegabriellijonalasinioladim1},{bertinisolegabriellijonalasinioladim2},
{bertinisolegabriellijonalasinioladim3}}
solved some of the same problems employing
the formalism of fluctuating hydrodynamics.
The results obtained for the probability functional of a given
density profile or for the integrated current distribution are
important since they provide
new insights into the properties of nonequilibrium steady states.
Thus it would be highly desirable to establish how general or universal
these results are.
In order to investigate this generality issue, in this paper,
we go beyond the specificity of the exclusion process, and study the effects of
varying the interactions between the particles as well as varying the microscopic
dynamics in boundary driven lattice gases.

Our main "technical" results are the derivation of
the weak interaction limits of both the nonequilibrium free energy and
the distribution of the integrated particle current. From the
explicit form of the free energy, we can then deduce the density profile and the
correlations (effective interactions) for various interactions (pair or triplet)
and various dynamics [zero-range~\cite{zrp-origin} and short-range
(misanthropic~\cite{Cocozza}) processes].

Our main "physical"  findings are that the density profiles are nonlinear and
they depend on both the interactions and the dynamics. Furthermore, the details
of the dynamics are found to change the
the correlations from short- to long-range, and they are shown to be able to
change the sign of the effective interactions. As to the
distributions of the integrated particle current, we verify that, for
all the interaction- and dynamics combinations we studied, they
are in agreement with the recently proposed addititvity principle which
allows to construct this distribution function from the knowledge of
its first two moments \cite{bodineauderrida}.

We start by introducing the models of interacting lattice gases and
deriving dynamical rules satisfying detailed balance in the absence of
boundary drive (Sec.\ref{Model}). Next we describe on the example of
free particles how the models can be formulated in terms of field
theories (Sec.\ref{Free}). Zero-range and short-range (misanthropic) processes
are worked out in Sections \ref{ZRP} and \ref{SRP}, respectively.
There we provide explicit expressions for the effective free energies.
Finally, Sec.\ref{JJJ} is devoted to a separate treatment
of the integrated current distribution.

\section{The model}\label{Model}
\subsection{Lattice gas with onsite interaction and boundary drive}
A $d=1$ dimensional lattice gas is considered with hopping dynamics
in the bulk and with particle injection and removal at the ends of the chain.
The state of the system $\vec n\equiv \{n_0,n_1,...,n_L\}$ is
specified by the number of particles $n_i=0,1,...,\infty$
at lattice sites $i$ ($i=0,1,...,L$), and
the interactions are assumed to be local
\begin{equation}
E(\vec n)=\sum_{i=0}^{L}h_i
\end{equation}
where $h_i=h(n_i)$ is the energy of interactions among particles at the same site.
We shall treat the simplest case
of pair interactions, $h(n)=\eps n (n-1)$. However triplet interactions,
$h(n)=\eps n (n-1)(n-2)$,
will also be considered occasionally either for demonstrating the effects of competing interactions
\begin{equation}\label{compete}
h_i=h(n_i)=\eps \left(\pm n_i(n_i-1)+
 \lambda n_i(n_i-1)(n_i-2) \right)\, .
\end{equation}
or for probing the universality of our results with respect to varying the type
of interaction. For $\lambda=0$, we restrict our study to the + sign in the
r.h.s. of (\ref{compete}) in order to have stability against
collapse of all particles on a single site, while for $\lambda > 0$ both signs
will be considered.

The dynamics of the system is described in terms
of a master equation for the
time evolution of the probability $P(\vec n,t)$ of a given particle configuration
\begin{equation}
\partial_t P(\vec n,t)={\cal L}_{D}P(\vec n,t)+{\cal L}_{BC}P(\vec n,t) \,
\end{equation}
where the first and second terms on the right-hand side represent the
nearest-neighbor hopping in the bulk and the injection-removal terms at the
boundaries. The bulk term is given through the rate of hopping $w_{i\rightarrow i\pm1}(\vec n)$
from site $i$ to $i\pm1$ in a state $\vec n$ as
\begin{eqnarray}
&&\hspace{-1.2cm}{\cal L}_{D}P(\vec n,t)=-\sum_{i=0}^{L-1}w_{i\rightarrow i+1}(\vec n)P(\vec n,t)-
			\sum_{i=1}^{L}w_{i\rightarrow i-1}(\vec n)P(\vec n,t)	       \\
&&+\sum_{i=0}^{L-1}w_{i\rightarrow i+1}(\vec n_{i+1\rightarrow i})P(\vec n_{i+1\rightarrow i},t)
+\sum_{i=1}^{L}w_{i\rightarrow i-1}(\vec n_{i-1\rightarrow i})P(\vec n_{i-1\rightarrow i},t)
\nonumber
\end{eqnarray}
where state $\vec n_{i\rightarrow j}$ is obtained from $\vec n$ by moving a particle from $i$ to $j$.
The injection and removal of particles at the boundaries $i=0$ and $i=L$
are described by the terms
\begin{eqnarray}
{\cal L}_{BC}P(\vec n,t)=&&-\left [\,w_0^+(\vec n)+w_L^+(\vec n)+
		       w_0^-(\vec n)+w_L^-(\vec n)\,\right ]\,P(\vec n,t)
			\\
&&+\,w_0^+(\vec {n}_{0^-})\,P(\vec n_{0^-},t)+w_0^-(\vec n_{0^+})\,P(\vec n_{0^+},t)
			\nonumber \\
&&+\,w_L^+(\vec n_{L^-})\,P(\vec n_{L^-},t)+w_L^-(\vec n_{L^+})\,P(\vec n_{L^+},t)
\nonumber
\end{eqnarray}
where $w_0^+(\vec n)$, $w_0^-(\vec n)$, $w_L^+(\vec n)$ and $w_L^-(\vec n)$ are the rates
of adding ($+$) or removing ($-$) a particle at site $0$ or $L$ in the state $\vec n$
and, furthermore, the state $\vec n_{i+(-)}$ is obtained from $\vec n$ by adding (removing)
a particle at site $i$.

\subsection{Choice of dynamics}
\label{choiceofdyn}
\subsubsection{Equilibrium distribution}

In order to specify the dynamics, we shall assume that if the
injection and removal rates are such that there is not
net flux (particle or energy current) through the system then
equilibrium is reached. Due to the onsite nature of the interaction, the
equilibrium distribution factorizes, it becomes the grand-canonical distribution
\begin{equation}
P_{\text{eq}}(\vec n)=\prod_i p_{\text{eq}}(n_i),\quad
p_{\text{eq}}(n_i)=\frac{1}{\Xi}\frac{\zeta^{n_i}\ee^{-h_i/T}}{n_i!}
\label{Peq}
\end{equation}
where $T$ and $\zeta$ are characterizing the temperature and fugacity
of the outside heat and particle bath, and $\Xi$ is a normalization factor. In the following,
we shall absorb $T$ in the interaction strength $\eps$ and the
high temperature limit studied below will simply mean that  $\eps$ is small.

\subsubsection{Hopping rates}

The bulk hopping rates may now be defined by making the natural (though not
necessary) assumption that they satisfy
detailed balance with $P_{\text{eq}}$, and that the detailed balance remains valid
in the presence of the boundary drive, as well. This assumption yields the following condition
for the rate of hopping to the right
\begin{eqnarray}
\frac{w_{i\rightarrow i+1}(\vec n)}{w_{i+1\rightarrow i}(\vec n_{i\rightarrow i+1})}&=&
\frac{n_i\ee^{-[E(\vec n_{i\rightarrow i+1})-E{(\vec n)]/2}}}
{(n_{i+1}+1)\ee^{-[E{(\vec n)}-E{(\vec n_{i\rightarrow i+1})]/2}}}\label{sr-rate}\\
&=&\frac{n_i\ee^{h(n_i)-h(n_i-1)}}{(n_{i+1}+1)\ee^{h(n_{i+1}+1)-h(n_{i+1})}}\label{zr-rate}
\, ,
\end{eqnarray}
and similar condition with $i+1$ replaced by $i-1$ applies for the rate of hopping to the left.
Clearly, the hopping rates are not uniquely determined by detailed balance.
The remaining arbitrariness is not relevant for equilibrium but,
in the presence of a drive,
the steady state properties do depend on details of the
dynamics. In order to
have some idea about the importance of various choices, we shall consider two
significantly different rates.

First, the rates will be chosen so that they
depend only on the energy change at the site from which the hopping originates. As can be
seen from (\ref{zr-rate}), a choice satisfying this condition is as follows
\begin{equation}
w_{i\rightarrow i+1}^{\rm (zr)}(\vec n)=Dn_i\ee^{h(n_i)-h(n_i-1)}\,
\label{zr-diff-rate}
\end{equation}
where $D$ is the "diffusion coefficient" setting the timescale.
Dynamics defined by the above rate is an example of the so called
{\em zero range processes} \cite{{zrp-origin},{zrp-review}}
where, in general, the rate is an arbitrary function of $n_i$. Thus, the first half
of our study can be viewed as an investigation of
zero range processes in the presence of a boundary drive.

For a second choice, a more standard hopping dynamics will be used with rates
depending not only on the energy change at the starting position of the hopping particle
but on the total energy change due to the hopping.
This choice follows from the eq.(\ref{sr-rate}) version of detailed balance
\begin{equation}
w_{i\rightarrow i+1}^{\rm (sr)}(\vec n)=Dn_i\ee^{-[h(n_i-1)+h(n_{i+1})-h(n_i)-h(n_{i+1})]/2}\,
\label{sr-diff-rate}
\end{equation}
where a subscript in $w^{\rm (sr)}$ signifies the name
{\it short range process} \cite{Cocozza}
we shall use in order to distinguish the resulting dynamics
from the zero range process. An important feature of the short range process is that
tuning the interactions $(0\le \eps < \infty, \lambda=0)$ so that one extrapolates
between the exactly solvable noninteracting limit $(\eps =0)$ and the much investigated
and also exactly solvable \cite{derridalebowitzspeer1,bertinisolegabriellijonalasinioladim3}
boundary driven symmetric exclusion process $(\eps \to \infty )$.

\subsubsection{Boundary drive (injection and removal rates)}
Nonequilibrium drive is introduced into the model through injection and removal of particles
at the ends ($j=0$ and $L$) of the chain. The rates can be set again by referring to
detailed balance condition with the baths attached to the ends
of the chain. The baths are assumed to contain noninteracting particles with the chemical
potentials set to produce the appropriate injection and removal rates.

For zero range
process, the removal can be imagined as diffusion into the baths. The rates are
independent of the properties of the baths and given by (\ref{zr-diff-rate}) with
$D$ replaced by new (externally controlled) parameters $\beta$ and $\gamma$, yielding
\begin{equation}
w^{(\rm zr)-}_0(\vec n)=\gamma n_0\ee^{h(n_0)-h(n_0-1)} \quad ; \quad
w^{(\rm zr)-}_L(\vec n)=\beta n_L\ee^{h(n_L)-h(n_L-1)} \, .
\label{zr-removal-rate}
\end{equation}
Within the picture of zero range process, the injections
(moving a particle from the bath to the end sites)
are independent of state of the end sites, and thus they are constants $\alpha$ and $\delta$
determined by the properties of the baths
\begin{equation}
w^{(\rm zr)+}_0(\vec n)=\alpha \quad ; \quad
w^{(\rm zr)+}_L(\vec n)=\delta \, .
\label{zr-injection-rate}
\end{equation}

The boundary rates for the short range process are determined along similar lines.
The rate of removal follows from (\ref{sr-rate}) and, since the particles in the
baths are not interacting, they are given by
\begin{equation}
w^{(\rm sr)-}_0(\vec n)=\gamma n_0\ee^{[h(n_0)-h(n_0-1)]/2} \quad ; \quad
w^{(\rm sr)-}_L(\vec n)=\beta n_L\ee^{[h(n_L)-h(n_L-1)]/2} \, .
\label{sr-removal-rate}
\end{equation}
In order to satisfy the detailed balance (\ref{sr-rate}), the injection rates must also depend
on the energy change due to the adding a particle and, using again that the particles
in the bath do not interact, we find
\begin{equation}
w^{(\rm sr)+}_0(\vec n)=\alpha \ee^{[h(n_0)-h(n_0+1)]/2} \quad ; \quad
w^{(\rm sr)+}_L(\vec n)=\delta \ee^{[h(n_L)-h(n_L+1)]/2} \, .
\label{sr-injection-rate}
\end{equation}

We have now finished the description of the two models we shall be
concerned with in this paper. Eqs.
(\ref{zr-diff-rate},\ref{zr-removal-rate},\ref{zr-injection-rate})
define the boundary driven zero range process while the rates given by eqs.
(\ref{sr-diff-rate},\ref{sr-removal-rate},\ref{sr-injection-rate})
define the corresponding short range process. As one can easily verify,
the case of
\begin{equation}
\frac{\alpha}{\gamma}=\frac{\delta}{\beta}
\end{equation}
corresponds to equilibrium with the baths (no drive) and the steady state
distribution is the equilibrium distribution given by eq.(\ref{Peq}).
Nonequilibrium drive is present when the above equality is violated.

\section{Preparing for the perturbation expansion: Noninteracting particles}
\label{Free}
\subsection{From the master equation to path integrals}
We shall now perform the mapping of the master equation for the probability
that the system is in
the occupation numbers configuration $\vec{n}$ at time $t$, namely,
$P(\vec{n};t)$, to a
field theory, where approximation techniques inherited from our experience
with equilibrium systems can be transferred without any difficulty.

We only briefly sketch the procedure (see \cite{mattisglasser} for a recent
review). We build up the state vector
$|\Psi(t)\rangle=\sum_{\vec{n}}P(\vec{n},t)|\vec{n}\rangle$ and rewrite the
master equation for $P$
in terms of and evolution equation for $|\Psi(t)\rangle$.
This equation can be cast in the form
\begin{equation}
\frac{\dd|\Psi(t)\rangle}{\dd t}=\hat{\cal L}_0|\Psi(t)\rangle
\end{equation}
where $\hat{\cal L}_0$ is an evolution operator acting on the
microstates $|\vec{n}\rangle$ indexed by the local particle
numbers. Not suprisingly, the evolution operator $\hat{\cal L}_0$ is
conveniently expressed in terms of
the creation and annihilation operators $a^{\dagger}_i, a_i$ related
to each local occupation
number $n_i$. For the rates specified in the previous subsection,
but at $\eps=0$, it takes the following form
\begin{equation}\begin{split}
-\hat{\cal L}_0=&D\sum_i\sum_{j\text{ nn of }i}
(a_i^\dagger-a_j^\dagger)a_i\\&+\alpha(1-a_0^\dagger)+\delta(1-a_L^\dagger)\\
&+\gamma(a_0^\dagger-1)a_0+\beta
(a_L^\dagger-1)a_L
\end{split}\end{equation}
One can of course recognize that at $\eps=0$ one is formally dealing with
a free boson hamiltonian.
The next step consists in converting the computation of expectation values
of physical observables
into the evaluation of a path-integral over two fields $\hat{a}_i(t)$ and $a_i(t)$.
This step leads
to averaging a physical observable $A(\{n_i\})$ in the following way:
\begin{equation}
\langle A(\vec{n})\rangle=\int{\cal D}\hat{a}_i{\cal D}a_i\;
\tilde{A}(\vec{a}(t))\;\ee^{-S_\eps[\hat{a},a]}
\end{equation}
where
\begin{equation}\begin{split}
S_0[\hat{a},a]=-\sum_i
a_i(t)+\int_0^t\dd\tau\Big[&\sum_i\Big(\hat{a}_i\p_\tau
a_i+D\sum_j(\hat{a}_i-\hat{a}_j)a_i\Big)\\&+\alpha(1-\hat{a}_0)+\gamma
(\hat{a}_0-1)a_0\\&+
\delta(1-\hat{a}_L)+\beta
(\hat{a}_L-1)a_L\Big]
\end{split}\end{equation}
Note that we have moved the initial condition to $t=-\infty$, in order to sit
directly in the
steady-state (initial condition independent). The prescription to obtain $\tilde{A}$
is as follows: normal order $A(\{a_i^\dagger a_i\})$ and then
replace the operators $a^\dagger_i$ and $a_i$ by the fields one and $a_i(t)$
respectively. This yields $\tilde{A}(\vec{a}(t))$.
We have adopted a path-integral formulation because of the large toolbox that
goes with it. Besides, it formally provides us with a sort of dynamical
partition function on which series expansion are straightforward.
The efficiency of the mapping lies in the following observation: at $\eps=0$,
that is for independent
particles, the Gaussian field theory exactly encodes Poissonian statistics.
Hence the high-temperature ($\eps\to 0$) expansion that will be performed next
consists in expanding around the
Poissonian distribution. For free particles, as expected, the stationary
measure factorizes as a
product of local Poissonian distributions,
\begin{equation}
P(\vec{n})=\prod_i\ee^{-\zeta_i}\frac{\zeta_i^{n_i}}{n_i!}
\label{Prod-state}
\end{equation}
Hence, for free particles, there is local equilibrium, and the density field
$\rho_i=\langle n_i\rangle$ is identical to the fugacity field $\zeta_i$. This means that the
properties of the system are the same as those that one would obtain in
equilibrium if one imposed a space dependent fugacity $\zeta_i$. Here the explicit
expression of the fugacity is given by, in the continuum limit, setting
$x=i/L\in[0,1]$,
\begin{equation}\label{freeprofile}
\zeta(x)=\zeta_0+(\zeta_1-\zeta_0)x+\frac{\zeta_0-\zeta_1}{\beta\gamma L}\left(\gamma
x-\beta(1-x)\right)+{\cal O}(L^{-2})
\end{equation}
where we have set $\zeta_0=\frac{\alpha}{\gamma}$ and
$\zeta_1=\frac{\delta}{\beta}$ as the fugacities of the reservoirs. The field theory described by $S_0$ is free, which results in $a(x,t)$ being a
nonfluctuating field set to its average expression $\rho(x)=\zeta(x)$. This means in particular that, for $\eps=0$,
\begin{equation}
\int{\cal D}\hat{a}_i{\cal D}a_i\tilde{A}(\{a_i(t)\})\ee^{-S_0}=\tilde{A}(\{\rho_i\})
\end{equation}
whatever the observable $A$.\\

Another central quantity is the response function of
the system to particle injection: the probability $G(x,y;t-t')$ that there is a particle at $x$ at time $t$ given
that there was one at $y$ at time $t'$ is given by
\begin{equation}
G(x,y;\tau)=\frac{2}{L}\sum_{n\in\mathbb{Z}}\sin(n\pi x)\sin(n\pi y)\ee^{-\pi^2 n^2 \tau/L^2}
\end{equation}
The time integrated response function $\hat{G}(x,y)=\int_0^\infty\dd\tau
G(x,y;\tau)$ will also be needed
\begin{equation}\begin{split}
\hat{G}(x,y)=&L\left[x(1-y)\Theta(y-x)+y(1-x)\Theta(x-y)\right]+\\
&\frac{1}{\gamma}(1-x)(1-y)-\frac{1}{\beta}x y\\&-\frac{1}{L\beta^2\gamma^2}\left(\gamma
x-\beta(1-x)\right)\left(\gamma
y-\beta(1-y)\right)+{\cal
O}(L^{-2})
\end{split}\end{equation}
In technical terms $G$ is simply the free propagator of the theory.\\
\subsection{Effective free energy in the steady-state}
The probability $P[\{n_i\}]$ to observe a given occupation number configuration
$\{n_i\}$ (or alternatively, a given proifle $n(x)$) is given by
\begin{equation}
P[\{n_i\}]=\langle\prod_i\delta(n_i-a_i^\dagger a_i)\rangle\end{equation}
We further define the effective free energy ${\cal F}[n]$ of the profile $n(x)$ by
\begin{equation}
{\cal F}[n]=-\lim_{L\to\infty}\frac{\ln P[n]}{L}
\end{equation}
To access ${\cal F}$ we first pass to the generating
function
\begin{equation}
\hat{P}[\{z_i\}]=\sum_{\{n_i\}}\prod_jz_j^{n_j}P[\{n_i\}]
\end{equation}
then work directly on
\begin{equation}
\Omega[\{z_i\}]=-\lim_{L\to\infty}\frac{\ln\hat{P}[\{z_i\}]}{L}
\end{equation}
which plays the r\^ole of an effective grand-potential. Our task is to compute
\begin{equation}
\hat{P}[\{z_i\}]=\langle\ee^{\sum_i(z_i-1) a_i(t)}\rangle
\end{equation}
where the brackets denote the weighted path-integral defined by the action $S_0$. Using again a
continuum limit notation, we find, as expected, that
\begin{equation}
\hat{P}[\{z(x)\}]\Big|_{\eps=0}=\ee^{L\int_0^1(z(x)-1) \rho(x)}
\end{equation}
from which we recover that the a steady-state probability distribution function is a product
of local Poissonian distributions:
\begin{equation}
P[\{n_i\}]=\prod_i\ee^{-\rho_i}\frac{\rho_i^{n_i}}{n_i!}
\end{equation}
Going to a continuum notation we find
\begin{equation}
{\cal F}[n]=\int_0^1\dd x\left[\rho(x)-n(x)+n(x)\ln\frac{n(x)}{\rho(x)}\right]
\end{equation}
in agreement with the mathematically precise construction of
the continuum limit \cite{derridalebowitzspeer2}.
At this stage, the present paragraph looks like a very technical rephrasing of simple
properties. We are now ready, however, to attack the case of interacting systems.

\section{Zero Range Process}
\label{ZRP}
\subsection{Evolution operator and field-theory}
We will restrict our analysis to the case of pair repulsion, namely
\begin{equation}
h(n)=\eps n (n-1)
\end{equation}
The master equation equation can be cast in the form
\begin{equation}
\frac{\dd|\Psi(t)\rangle}{\dd t}=\hat{\cal L}_\eps|\Psi(t)\rangle
\end{equation}
where for the rates specified in (\ref{zr-diff-rate},\ref{zr-removal-rate},\ref{zr-injection-rate}), the evolution operator $\hat{\cal L}_\eps$ takes the following form
\begin{equation}\begin{split}
-\hat{\cal L}_\eps=&D\sum_i\sum_{j\text{ nn of }i}
(a_i^\dagger-a_j^\dagger)\ee^{2\eps a_i^\dagger
a_i}a_i\\&+\alpha(1-a_0^\dagger)+\delta(1-a_L^\dagger)\\&+\gamma(a_0^\dagger-1)\ee^{2\eps a_0^\dagger a_0}a_0+\beta
(a_L^\dagger-1)\ee^{2\eps a_L^\dagger a_L}a_L
\end{split}\end{equation}
One can of course recognize that at $\eps=0$ one recovers the free boson hamiltonian of the
previous section. The corresponding action reads
\begin{equation}\begin{split}
S_\eps[\hat{a},a]=-\sum_i
a_i(t)+\int_0^t\dd\tau\Big[&\sum_i\Big(\hat{a}_i\p_\tau
a_i+D\sum_j(\hat{a}_i-\hat{a}_j)\ee^{(\ee^{2\eps}-1)\hat{a}_i
a_i}a_i\Big)\\&+\alpha(1-\hat{a}_0)+\gamma
(\hat{a}_0-1)\ee^{(\ee^{2\eps}-1)\hat{a}_0a_0}a_0\\&+
\delta(1-\hat{a}_L)+\beta
(\hat{a}_L-1)\ee^{(\ee^{2\eps}-1)\hat{a}_L a_L}a_L\Big]
\end{split}\end{equation}
Our calculations will be based on an expansion of $S_\eps$ to first order in $\eps$:
\begin{equation}\label{actionZRP}
S_\eps=S_0+2\eps \int\dd
t\Bigg[\sum_{i,j}(\hat{a}_j-\hat{a}_i)\hat{a}_ia_i^2+\gamma(\hat{a}_0-1)\hat{a}_0a_0^2+\beta(\hat{a}_L-1)\hat{a}_La_L^2\Bigg]
+{\cal O}(\eps^2)\end{equation}
\subsection{Effective free energy, profile and correlations}
In order to evaluate
\begin{equation}
\hat{P}[z(x)]=\langle\ee^{L\int_0^1\dd x(z(x)-1)a(x,t)}\rangle
\end{equation}
we rely on a cumulant expansion. This form is particularly well-suited for a cumulant expansion,
which we write
as
\begin{equation}
\hat{P}[z(x)]=\langle\exp\left[L\sum_{n\geq 1}\frac{1}{n!}\int_0^1\prod_{j=1}^n (z(x_j)-1)
W^{(n)}(x_1,...,x_n)+\right]
\end{equation}
where $W^{(n)}$ is the $n$-point connected correlation function of the field
$a$ at equal times, in the steady state.
In practice it is convenient to introduce the auxiliary fields
$\bar{\phi}=\hat{a}-1$ and $\phi=a-{\zeta}$. Note that
\begin{equation}
{\rho}(x)=\langle n(x)\rangle=\langle a(x)\rangle=W^{(1)}(x)
\end{equation}
is so far undetermined. It will be useful to split $W^{(2)}$
into two pieces, denoted by $W^{(2)}_{\text{loc}}$ and $W^{(2)}_{\text{NE}}$,
corresponding to the local delta correlated piece in $W^{(2)}$ (namely the
diagonal part) and the genuinely nonequilibrium contribution, respectively. In order to obtain $W^{(2)}$ one first directly computes the field connected
correlation function to leading order in $\eps$:
\begin{equation}\begin{split}
\langle\phi(x_1,t_1)\phi(x_2,t_2)\rangle_c=&
\frac{2\eps}{L}\int_0^\infty \dd t
\int_0^1\dd y\Big( \zeta(y)^2(\p_y^2G(x_1,y;t_1-t)G(x_2,y;t_2-t)\\&+G(x_1,y;t_1-t)\p_y^2G(x_2,y;t_2-t) )\Big)
\end{split}\end{equation}
where the usual contractions between a bared and a non-bared field were carried out. This
yields,
\begin{equation}\label{corrfieldZRP}
W^{(2)}(x,y)=W_{\text{loc}}^{(2)}(x,y)=-2\eps\zeta^2(x)\delta(L(x-y))
\end{equation}
as if local equilibrium would hold (to this order in $\eps$ at least).
Note that the time depependent correlations are obtained by the same token.
The profile for the ZRP reads
\begin{equation}\label{profZRP}\begin{split}
{\rho}(x)=\zeta(x)-2\eps\zeta(x)^2
\end{split}\end{equation}
Let us compare with what we would obtain in equilibrium for the distribution
(\ref{Peq}). The density-fugacity relationship and the local particle
number fluctuations would read
\begin{equation}
\rho=\langle n \rangle=\zeta-2\eps \zeta^2,\;\;\langle n^2\rangle_c=\zeta-4\eps
\zeta^2
\end{equation}
A quick glance at (\ref{profZRP},\ref{corrfieldZRP}) shows that the zero range
process, to first order in $\eps$ and to leading order in $L$ at least, appears
to be consistent with local equilibrium. The existence of local equilibrium
can actually be proved in general, independently of the explicit
form of $h(n)$ provided the transition rates are given by the expressions
(\ref{zr-diff-rate},\ref{zr-removal-rate},\ref{zr-injection-rate}).
At the field theory level, it follows by inspection of the
corresponding Feynman diagrams which indicates that
local equilibrium holds to all orders in $\eps$
due to the fact that the interaction arising from diffusion
in the bulk is proportional to the Laplacian of the response field.
Hence, by Wick's theorem, the interactions are proportional to the
Laplacian of the propagator $G$ which is a delta function in space.
Thus spatial correlations cannot be built up and the steady state
measure remains a product measure just as is the case for
noninteracting particles (\ref{Prod-state}). It should be noted,
that the existence of local equilibrium can be more simply proved \cite{Gunter}
by just substituting product measure into the steady state
master equation and deriving an equation for the local fugacity
(\ref{freeprofile}).\\

For zero range processes the effective free energy can thus directly be
obtained from the equilibrium distribution (\ref{Peq}) in which one has
substituted the local fugacity by its expression in terms of the local average
density.

\section{Short range process}
\label{SRP}
\subsection{Evolution operator and action}
As announced in \ref{choiceofdyn}, we now wish to study a different set of dynamical rules, those of the
short-range process. For definiteness, we confine the present paragraph to the case of
on-site pair repulsion, namely, $h(n)=\eps n(n-1)$. One of the reasons for this
choice is that, part from the $\eps=0$ limit which reduces to free particles,
the $\eps\to\infty$ limit exactly corresponds to the Symmetric Exclusion
Process (SEP). Given that the scaling variable that appear in our expansions is
$\eps\rho$, we may have the hope to connect our results with the low density
behavior of the results obtained for the SEP. Again, for this dynamics, it is possible to
write the master equation in the form of an imaginary time Schr\"odinger equation with an
evolution operator
\begin{equation}\begin{split}
-\hat{\cal L}_\eps=&D\sum_i\sum_{j\text{ nn of }i}
(a_i^\dagger-a_j^\dagger)\ee^{\eps\hat{n}_i-\eps\hat{n}_j}a_i\\&+\alpha(1-a_0^\dagger)\ee^{-\eps\hat{n}_0}+\delta(1-a_L^\dagger)\ee^{-\eps\hat{n}_L}\\
&+\gamma(a_0^\dagger-1)\ee^{\eps\hat{n}_0}a_0+\beta
(a_L^\dagger-1)\ee^{\eps\hat{n}_L}a_L
\end{split}\end{equation}
In practice our analysis will be limited to the first nontrivial order in
$\eps$, and the corresponding action reads
\begin{equation}\label{actionSRP}\begin{split}
S_\eps=&S_0+\eps \int\dd
t\Bigg[\sum_{i,j}(\hat{a}_j-\hat{a}_i)(\hat{a}_ia_i-\hat{a}_j a_j)\\&-\alpha
(1-\hat{a}_0)\hat{a}_0
a_0+\gamma(\hat{a}_0-1)\hat{a}_0a_0^2\\&-\delta(1-\hat{a}_L)\hat{a}_L
a_L+\beta(\hat{a}_L-1)\hat{a}_La_L^2\Bigg]
\end{split}\end{equation}
The expanded action (\ref{actionSRP}) will be the starting point of our computations.

\subsection{Effective free energy for onsite pair repulsion}
As explained earlier, we shall focus on the generating function of the
probability distribution,
\begin{equation}
\hat{P}[z(x)]=\langle\ee^{L\int_0^1\dd x (z(x)-1) a(x,t)}\rangle
\end{equation}
Note that if one would start from a hamiltonian of the
form
\begin{equation}
h(n)=\eps\sum_{p\geq 2} g_p n(n-1)...(n-p+1)
\end{equation}
where the $g_p$'s are order one constants, then the cumulant expansion would
require to go as far as $W^{(p)}$, to leading order in $\eps$. Of course, at
finite $\eps$, all cumulants are needed. In the present case of onsite pair
repulsion, namely with
$
h(n)=\eps n(n-1)
$
we are thus left with the computation of the first and second cumulant of
$a(x,t)$. We find that ${\rho}(x)=\langle a(x,t)\rangle$ has the following
expression:
\begin{equation}\begin{split}
{\rho}(x)=&\zeta(x)-2\eps\gamma
(\zeta_0+\zeta(0))\zeta(0)\hat{G}(x,0)
-2\eps\beta(\zeta_1+\zeta(1))\zeta(1)\hat{G}(x,1)\\
&-\frac{\eps}{L}\int_0^1\dd y\;\zeta(y)\p_y\zeta(y)\p_y \hat{G}(x,y)
\\=&(\zeta_0-2\eps\zeta_0^2)(1-x)+(\zeta_1-2\eps \zeta_1^2)x
+\eps(\zeta_1-\zeta_0)^2x(1-x)\\&+\frac{\zeta_0-\zeta_1}{\beta\gamma L}\Bigg(\gamma
x\Big(1-\eps[\zeta_0(3-2x)+\zeta_1(2+2x)]\Big)\\&\qquad-\beta(1-x)\Big(1-\eps[\zeta_1(1+2x)+\zeta_0(4-2x)]\Big)\Bigg)+{\cal
O}(L^{-2})
\end{split}\end{equation}
It may be seen that
\begin{equation}
{\rho}(x)=\zeta(x)-2\eps\zeta(x)^2-\eps(\zeta_0-\zeta_1)^2 x(1-x)+{\cal O}(L^{-1})
\end{equation}
thus showing that there is no local equilibrium in the short range process.
The density-density correlation function $C^{(2)}(x,y)=\langle n(x)n(y)\rangle_c$ takes the simple form
\begin{equation}\begin{split}
C^{(2)}(x,y)=(\zeta(x)-2\eps\zeta(x)^2)\delta(L(x-y))+W^{(2)}(x,y)
\end{split}\end{equation}
with
\begin{equation}\label{W2pairs}\begin{split}
W^{(2)}(x,y)=&-2\eps\zeta(x)^2\delta(L(x-y))
\\&-\frac{\eps}{L}(\zeta_0-\zeta_1)^2\left[x(1-y)\Theta(y-x)+y(1-x)\Theta(x-y)\right]
\end{split}\end{equation}
which now features a nonzero long range $W^{(2)}_{\text{NE}}$ piece:
\begin{equation}
\begin{split}
W_{\text{NE}}^{(2)}(x,y)=&-\frac{\eps}{L}(\zeta_0-\zeta_1)^2\left[x(1-y)\Theta(y-x)+y(1-x)\Theta(x-y)\right]
\end{split}\end{equation}
It is possible to return to the effective free energy
\begin{equation}\label{freeenergypairs}\begin{split}
{\cal F}[n(x)]=&\int_0^1\dd x
\left[{\rho}(x)+\eps\rho(x)^2-n(x)+n(x)\ln\frac{n(x)}{{\rho}(x)+2\eps\rho(x)^2}+\eps
n(x)(n(x)-1)\right]\\&-\frac 12\int_0^1\dd x\dd
y\left(\frac{n(x)}{{\rho}(x)}-1\right)W_{\text{NE}}^{(2)}(x,y)\left(\frac{n(y)}{{\rho}(y)}-1\right)
\end{split}\end{equation}
The first brackets on the rhs of (\ref{freeenergypairs}) corresponds to a system
in local equilibrium with respect to an effective fugacity $\rho+2\eps\rho^2$;
it already appeared for the zero range process. The integral in the second line illustrates the nonlocal
long-range nature
of the effective interactions in a nonequilibrium steady-state
($W^{(2)}_{\text{NE}}$ being negative, the corresponding contribution is
positive, which expresses the repulsive nature of the effective interactions). The fluctuations
of the total number of particles read
\begin{equation}\label{fluctSRP}
\Delta N^2=\Delta N^2_{\text{loc. eq.}}-L\frac{\eps}{12}(\rho_0-\rho_1)^2
\end{equation}
where $\Delta N^2_{\text{loc. eq.}}$ denotes the fluctuations of a
systems in local thermal equilibrium with the same fugacity profile (such as the
ZRP). The decrease of the global fluctuations is yet another consequence of the
development of long-range anticorrelations. As a coincidence, note that setting
in (\ref{fluctSRP}) $\eps=1$ yields exactly the symmetric exclusion process
expression for this quantity~\cite{derridalebowitzspeer2}.

\subsection{Effective free energy with onsite triplet repulsion}
\begin{figure}
$$\input{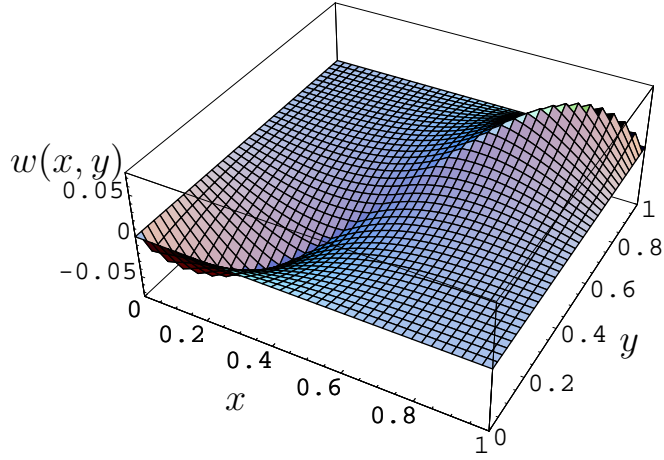}$$
\caption{Plot of $w(x,y)$ [see eq.(\ref{wxy})] as a function of $x,y\in[0,1]$.}\label{fig-w}
\end{figure}

Now we wish to explore the effect of varying the type of interaction by
studying the case
\begin{equation}
h(n)=\eps n (n-1)(n-2)
\end{equation}
which assigns an energy cost to the site proportional to the number
of triplets of particles are
present at the site. We find that the profile is given by
\begin{equation}
\rho(x)=\underbrace{\zeta(x)-3\eps\zeta(x)^3}_{\text{local
eq.}}-\eps x(1-x)(\zeta_0-\zeta_1)^2(\zeta_0+\zeta_1+\zeta_0(1-x)+\zeta_1 x)
\end{equation}
while the correlation function of the field reads
\begin{equation}\label{W2triplets}\begin{split}
W^{(2)}(x,y)=&-6\eps\zeta(x)^3\delta(L(x-y))
\\&-\frac{12\eps}{L}(\zeta_0-\zeta_1)^2(\zeta_0+\zeta_1)
\left[x(1-y)\Theta(y-x)+y(1-x)\Theta(x-y)\right]\\&
+\frac{12\eps}{L}(\zeta_0-\zeta_1)^3 w(x,y)
\end{split}\end{equation}
In (\ref{W2triplets}) the extra function $w(x,y)$ has the explicit expression
\begin{equation}
w(x,y)=\frac{16}{\pi^4}\sum^\prime_{n,m\geq 1}\frac{1}{m^2+n^2}
\left[\frac{1}{(m+n)^2}-\frac{1}{(m-n)^2}\right]\sin(n\pi x)\sin (m\pi y)
\label{wxy}
\end{equation}
where $\displaystyle{\sum^\prime_{n,m\geq 1}}$ means a summation over $n$ and $m$ of opposite parities. We have not been
able to come up with a closed expression for $w$ (though one might exist). The function $w$
vanishes at $(1/2,1/2)$ and does not contribute to the number of particles
fluctuations. Figure (\ref{fig-w})
shows a two-dimensional plot of $w$ which reveals that $w(x,y)$ conveys
anticorrelations for $x,y$ in the vicinity of the reservoir with the highest
density, while positive correlations develop close to the reservoir imposing
the lowest density.
The function $w$ stands for the deviation of the long-range component of the
density correlation function with respect to the pair interaction case (whether
at small $\eps$ as in (\ref{W2pairs}), or at $\eps\to\infty$ as computed by
Spohn~\cite{spohn}). It is worth stressing that, to our knowledge, no
microscopic model had, up to now, shown such strong deviations.\\ 

The fluctuations
of the total number of particles now read
\begin{equation}
\Delta N^2=\Delta N^2_{\text{loc. eq.}}-2L\eps(\zeta_0-\zeta_1)^2(\zeta_0+\zeta_1) 
\end{equation}
where $\Delta N^2_{\text{loc. eq.}}$ denotes the fluctuations of a
systems in local thermal equilibrium with the same fugacity profile (such as the
ZRP). In the present case the third cumulant is of order $\eps$ as well, and it
has the formal expression
and that 
\begin{equation}\label{W3triplet}\begin{split}
&W^{(3)}(x_1,x_2,x_3)=-6\eps\zeta(x_1)^3\delta(L(x_1-x_2))\delta(L(x_2-x_3))
\\&-\frac{6\eps}{L}\int\dd\tau\dd z(\zeta_0(1-z)+\zeta_1 z)G(x_1,z;\tau)G(x_2,z;\tau)G(x_3,z;\tau)
\end{split}\end{equation}
Denoting by $W^{(3)}_{\text{NE}}$ the function appearing in
the second line of the r.h.s. of Eq.~(\ref{W3triplet}) allows to write the
effective free energy in the form
\begin{equation}\label{freeenergytriplets}\begin{split}
{\cal F}[n(x)]&=\int_0^1\dd x
\left[{\rho}+\eps\rho^3-n+n\ln\frac{n}{{\rho}+3\eps\rho^3}+\eps
n(n-1)(n-2)\right]\\&-\frac 12\int_0^1\dd x\dd
y\left[\frac{n(x)}{{\rho}(x)}-1\right]
W_{\text{NE}}^{(2)}(x,y)\left[\frac{n(y)}{{\rho}(y)}-1\right]\\&
-\frac{1}{3!}\int_0^1\dd x\dd y\dd z
\left(\frac{n(x)}{{\rho}(x)}-1\right)
\left(\frac{n(y)}{{\rho}(y)}-1\right)
\left(\frac{n(z)}{{\rho}(z)}-1\right)
W_{\text{NE}}^{(3)}(x,y,z)
\end{split}\end{equation}
As one can see, the triplet repulsion generates not only two-body
long ranged repulsive interactions, but also three body interactions, repulsive
as well. This establishes a correspondence between the microscopics (the precise
form of $h(n)$) and the effective interactions generated by the boundary drive.

\subsection{Competing interactions}
Suppose we now focus on a local interaction energy of the form
\begin{equation}
h(n)=\eps\left[- n (n-1)+\lambda n (n-1) (n-2)\right],\;\lambda >0
\end{equation}
where the first term on the r.h.s., which favors pair condensation, is competing
with the second term on the r.h.s. which assigns a relative energy cost
$\lambda$ to the piling up of three or more particles. To first order in
$\eps$ the density correlation function is merely a linear combination of that
obtained for the pair and triplet interactions, namely
\begin{equation}\begin{split}
C^{(2)}(x<y)=\frac{\eps}{L}(\zeta_0-\zeta_1)^2&\Big[
(1-12\lambda(\zeta_0+\zeta_1))x(1-y)+12\lambda(\zeta_0-\zeta_1) w(x,y)\Big]
\end{split}\end{equation}
We can clearly see that, at sufficiently low densities, positive correlations
develop, thus providing a microscopic counterexample to the general belief
that  the driving of a current builds up anticorrelations (note, however, that
Spohn~\cite{spohn} had left this possibility {\it a priori} open).

\section{Integrated current distribution}
\label{JJJ}
Let $Q(t)$ be the net number of particles which have jumped from the left
reservoir into the system over the time interval $[0,t]$ (counted positively for
an actual jump from the reservoir into the system, and negatively for a jump out
of the system to the reservoir). The physical motivations for studying the
properties of $Q$ are detailed by Lebowitz and
Spohn~\cite{lebowitzspohn} who showed that this quantity, defined for a Markov
process (our boundary driven lattice gas with stochastic dynamics) plays a role analogous to
the phase space contraction rate for dynamical systems (and for which
Gallavotti and Cohen~\cite{gallavotticohen} proved their fluctuation theorem).
Below, we shall concerned with the calculation of the distribution of $Q(t)$.

\subsection{Free particles}
\begin{figure}
$$\input{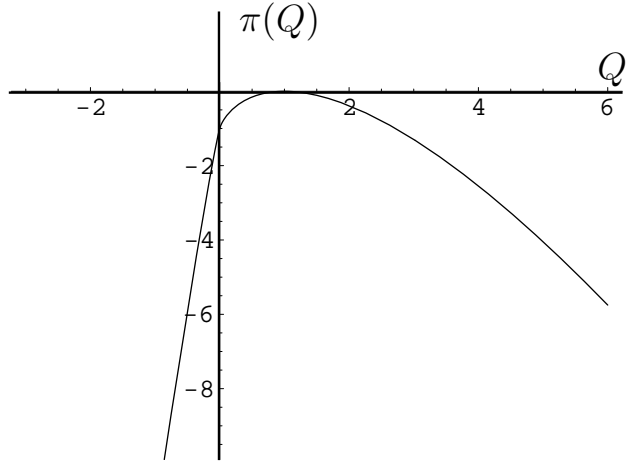}$$
\caption{Plot of $\pi(Q)$ in eq.(\ref{piQeq})
for free particles with $\rho_0=1\gg\rho_1$. Note the
strong asymmetry of $\pi$.}
\label{fig-piQ}
\end{figure}

Following the procedure described in \cite{derridalebowitz,derridadoucotroche},
we construct a master equation for
$P(\{n_i\},Q,t)$, the probability that the system is in state $\{n_i\}$ and with
$Q(t)=Q$ by explicitly separating those moves in phase space that increase $Q$
by one, decrease it by one, or leave it unchanged. Introducing a state vector
$|\psi(Q,t)\rangle=\sum_{\{n_i\}}P(\{n_i\},Q,t)|\{n_i\}\rangle$, we are left with an evolution
equation of the form
\begin{equation}
\frac{\dd|\psi(Q,t)\rangle}{\dd t}=-(\hat{H}_{1}+\hat{H}_{-1}+\hat{H}_0)|\psi(Q,t)\rangle
\end{equation}
where the operators $\hat{H}_\pm 1$ increase/decrease $Q$ by one, and where
$\hat{H}_0$ leave $Q$ unchanged. We are interested in the distribution function
of $Q(t)$ in the steady state, over large time intervals,
\begin{equation}
p(Q,t)=\langle \bp|\psi(Q,t)\rangle
\end{equation}
or, more conveniently, its generating function, namely,
\begin{equation}
\hat{p}(z,t)=\sum_{Q=-\infty}^{+\infty}z^Qp(Q,t)
\end{equation}
It turns out that the generating function can conveniently be expressed in terms of a path-integral
\begin{equation}
\hat{p}(z,t)=\int{\cal D}\hat{a}_i{\cal D}a_i\; \ee^{-S_{0,z}[\hat{a},a]}
\end{equation} 
where, for noninteracting particles
\begin{equation}\begin{split}
S_{0,z}[\hat{a},a]=-\sum_i
a_i(t)+\int_0^t\dd\tau\Big[&\sum_i\Big(\hat{a}_i\p_\tau
a_i+D\sum_j(\hat{a}_i-\hat{a}_j)a_i\Big)\\&+\alpha(1-z\hat{a}_0)+\gamma
(\hat{a}_0-z^{-1})a_0\\&+
\delta(1-\hat{a}_L)+\beta
(\hat{a}_L-1)a_L\Big]
\end{split}\end{equation} 
The action $S_{0,z}$ does not describe a stochastic process (unless $z=1$ for
which it reduces to $S_0$). However, it may be
readily seen that
\begin{equation}
\hat{p}(z,t)=\langle 
\ee^{\int_0^t\dd t (\alpha (z-1)-\gamma(1-z^{-1})a_0(t)) }\rangle_z
\end{equation}
where the brackets $\langle..\rangle_z$ now denote an average with respect to the 
process governed by $S_0$ in which $\alpha$ is formally replaced with $\alpha z$. Using that $a_0(t)$, for free
particles, is a nonfluctuating field with expression taken from (\ref{freeprofile}) by changing
$\alpha$ into $\alpha z$,
\begin{equation}
a_0(t)=\zeta_0 z-\frac{z\zeta_0-\zeta_1}{\gamma L}\left[1-\eps(4\zeta_0 z+\zeta_1)\right]
\end{equation}
we find that, in the long time limit, 
\begin{equation}
\lim_{t\to\infty}\frac{\ln\hat{p}(z,t)}{t}=\mu(z)=\frac{1}{L}\frac{z-1}{z}(\zeta_0 z-\zeta_1)
\end{equation}
It is {\it a posteriori} clear why subleading finite size corrections had to be
kept all along. The function $\mu(z)$ is the generating function for the cumulants of $Q$.
It is worth commenting on two
remarkable, yet expected, properties of $\mu(\zeta_0,\zeta_1,z)$. Namely, if we
exchange the roles of the
reservoirs, the distribution of $Q$ is turned into that of $-Q$, hence
\begin{equation}\label{P}
\mu(\zeta_0,\zeta_1,z)=\mu(\zeta_1,\zeta_0,z^{-1})
\end{equation}
But it also
satisfies the Gallavotti--Cohen property (see \cite{lebowitzspohn,derridadoucotroche} for a readable proof), namely
\begin{equation}\label{GC}
\mu(\zeta_0,\zeta_1,z)=\mu(\zeta_0,\zeta_1,\frac{\zeta_1}{\zeta_0 z})
\end{equation}
which is best known when rephrased as follows. Let $\pi(Q)$ be the large
deviation function related to $p(Q,t)$:
\begin{equation}
\pi(Q)=\lim_{t\to\infty}\frac{\ln p(Q,t)}{t}
\end{equation}
Besides, $\pi(Q)$ appears as the Legendre transform of $\mu(z)$ with respect to
 $\ln z$:
\begin{equation}
\pi(Q)=\text{max}_z\{\mu(\zeta_0,\zeta_1,z)-\ln z Q\}
\end{equation}
hence
\begin{equation}
\lim_{t\to\infty}\frac{1}{t}\ln\frac{p(Q,t)}{p(-Q,t)}=\pi(Q)-\pi(-Q)=\ln\frac{\zeta_0}{\zeta_1}Q
\end{equation}
Of course this can {\it a posteriori} be verified on the explicit expression of $\pi(Q)$:
\begin{equation}\begin{split}
\pi(Q)=&\frac{Q}{2}+\sqrt{\frac{Q^2}{4}+\zeta_0\zeta_1}
+\frac{\zeta_0\zeta_1}{\frac{Q}{2}+\sqrt{\frac{Q^2}{4}+\zeta_0\zeta_1}}\\&-(\zeta_0+\zeta_1)
-Q\ln\left[\frac{\sqrt{\frac{Q^2}{4}+\zeta_0\zeta_1}+\frac{Q}{2}}{\zeta_0}\right]
\end{split}
\label{piQeq}
\end{equation}
which is plotted on Fig.~(\ref{fig-piQ}).
The Gallavotti-Cohen theorem was shown~\cite{kurchan,lebowitzspohn} to hold under quite general conditions
for nonequilibrium steady-states resulting from Markovian dynamics and we shall
further comment upon it in subsection \ref{discaddprinc}.
\subsection{Short Range Process with pair repulsion}
We now repeat the strategy outlined in the previous paragraph, but for the
interacting case, with the short range process dynamics. We split the evolution operator $\hat{H}_\eps$ into three
pieces, each describing moves that increase, decrease of leave $Q$ unchanged.
Then we pass to a path integral in terms of which we find
\begin{equation}
\hat{p}(z,t)=\int{\cal D}\hat{a}_i{\cal D}a_i\; \ee^{-S_{\eps,z}[\hat{a},a]}
\end{equation} 

Having found $S_{\eps,z}$ we bring the calculation of $\hat{p}$ to that of a given observable with
respect to the original stochastic process in which $\alpha$ is replaced with
$\alpha z$:
\begin{equation}\begin{split}
\hat{p}(z,t)=\langle\exp\Big(-\int_0^t\dd
t\Big[\alpha(1-z)(1-\eps\hat{a}_0a_0)+\gamma(1-z^{-1})a_0(1+\eps\hat{a}_0a_0)\Big]\Big)\rangle_z
\end{split}\end{equation}
In order to evaluate the latter expectation value to leading order in $\eps$ we rely on a cumulant
expansion. 
Introducing the variable 
\begin{equation}
\omega(\zeta_0,\zeta_1,z)=\frac{z-1}{z}\left[(\zeta_0-\eps\zeta_0^2)z-(\zeta_1-\eps \zeta_1^2)-\eps (z-1)\zeta_0\zeta_1\right]
\end{equation}
we find that
\begin{equation}
L\mu=\omega-\frac{\eps}{3}\omega^2
\end{equation}
It is remarkable that the intermediate variable $\omega$ itself satisfies the two
invariance properties that $\mu$ is known to fulfill. Namely, the function
$\omega(\rho_0,\rho_1,z)$ verifies the left-right symmetry
\begin{equation}
\omega(\zeta_0,\zeta_1,z)=\omega(\zeta_1,\zeta_0,z^{-1})
\end{equation}
and the Gallavotti--Cohen property
\begin{equation}
\omega(\zeta_0,\zeta_1,z)=\omega(\zeta_0,\zeta_1,\frac{\zeta_1}{\zeta_0 z})
\end{equation}
That $\omega$ verifies (\ref{P},\ref{GC}) implies directly that $\mu$ possesses
the same invariance properties, as already identified in the SEP~\cite{derridadoucotroche}. One invariance property that our result does not possess, in
constrast to \cite{derridadoucotroche}, is the particle--hole symmetry.\\

The effective small parameter of the expansion is $\eps\rho$, where $\rho$ is the typical
density, hence our small $\eps$ expansion is seen to coincide with a low density expansion of the exact
result obtained at $\eps\to\infty$ by \cite{derridadoucotroche},
\begin{equation}
L\mu=\omega-\frac13
\omega^2+\frac{8}{45}\omega^3+{\cal O}(\omega^4)
\end{equation}
with
\begin{equation}
\omega=\frac{z-1}{z}\left[\frac{\zeta_0}{1+\zeta_0}z-\frac{\zeta_1}{1+\zeta_1}-
(z-1)\frac{\zeta_0}{1+\zeta_0}\frac{\zeta_1}{1+\zeta_1}\right]
\end{equation}
A natural question that arises next is how universal the result  obtained in the
exact calculation \cite{derridadoucotroche} is? How sensitive is it to
varying the microscopic interactions. In order to sort out this issue we have
performed a similar analysis for repulsive triplet interactions, with short
range dynamics.

\subsection{Short Range Process with triplet repulsion}
In the  same spirit as in the previous paragraph, we express the
generating function $\hat{p}(z)$
as the expectation value of an exponential observable with
respect to the lattice gas measure in
which $\alpha$ is replaced with $\alpha z$. 
We rely again on a cumulant expansion, and after a rather tedious
calculation, we arrive at the
following results. We now define the auxiliary variable $\omega$ by
\begin{equation}\begin{split}
\omega(\zeta_0,\zeta_1,z)=&\frac{z-1}{z}\left[(\zeta_0-\eps\zeta_0^3)z-(\zeta_1-\eps \zeta_1^3)-\eps
(z-1)\zeta_0\zeta_1(\zeta_0+\zeta_1)\right]\\&\times\left[1-\frac{\eps}{2}\frac{z-1}{z}(\zeta_0 z-\zeta_1)(\zeta_0+\zeta_1)\right]
\end{split}\end{equation}
which also obeys (\ref{P},\ref{GC}). We find that
\begin{equation}
L\mu=\omega-\frac{\eps}{10}\omega^3+{\cal O}(\eps^2)
\end{equation}
This expression unambiguously points at a different distribution function for the
integrated current. It further allows to connect the microscopics --the triplet
repulsion-- with the final form of $\mu$. A $p$-body interaction would yield a 
first correction to $\mu$ of the form $\omega^p$. Unfortunately we have not been
able to come by a physical interpretation for the intermediate quantity $\omega$.
\subsection{Zero Range Process with pair repulsion}
Finally we examine the case of the zero range process with pair repulsion, for
which we know that the steady state distribution follows local equilibrium.
It is then sufficient to start from the equilibrium expression for the free process
in which one has subsituted the current with its appropriate expression. It is
not hard to see, through a direct evaluation, that
\begin{equation}\begin{split}
\frac{\langle Q\rangle}{t}=\alpha-\gamma\langle
n_0\ee^{2\eps(n_0-1)}\rangle=\frac{(\zeta_0-2\eps\zeta_0^2)
-(\zeta_1-2\eps\zeta_1^2)}{L}\simeq\frac{\rho_0-\rho_1}{L}
\end{split}\end{equation}
which leads us to
\begin{equation}\label{muZRP}\begin{split}
L\mu(\zeta_0,\zeta_1,z)=
\frac{z-1}{z}\left[(\zeta_0-2\eps\zeta_0^2) z-(\zeta_1-2\eps\zeta_1^2)\right]
\end{split}\end{equation}
This form is identical to that predicted by Bodineau and
Derrida~\cite{bodineauderrida}) for zero range processes.
The Gallavotti-Cohen relation, in our particular case, now takes the form
\begin{equation}\begin{split}
\mu(\zeta_0,\zeta_1,z)=\mu(\zeta_0,\zeta_1,\frac{\zeta_0-2\eps\zeta_0^2}{\zeta_1-2\eps\zeta_1^2}\frac{1}{z})
\end{split}\end{equation}
It is remarkable that at equilibrium, that is for $\rho_0=\rho_1$, the current fluctuations $\langle Q^2\rangle$ is the
same for the ZRP and the SRP, while higher order cumulants (that is the whole distribution) are
different.

\subsection{Illustration of the additivity principle}\label{discaddprinc}
To conlude this section, we would like to illustrate how our explicit results
fit into the general framework provided by Bodineau and
Derrida~\cite{bodineauderrida}. In particular, they have deduced, from a postulated additivity principle, a
general expression for the integrated current distribution, from the sole
knowledge of $\langle Q\rangle(\rho_0,\rho_1)$ and $\langle
Q^2\rangle_c(\rho_0,\rho_1)$. One of the consequences of their findings is a
direct computation of the analog, in the Gallavotti-Cohen theorem, of the
"entropy production rate" (as formally defined by Lebowitz and
Spohn~\cite{lebowitzspohn}) in these
driven stochastic lattice gases. Defining, as in \cite{bodineauderrida},
\begin{equation}
D(\rho)=\frac{1}{t}\frac{\p}{\p \rho_0}\langle
Q\rangle(\rho_0,\rho)\Big|_{\rho_0=\rho},\;\sigma(\rho)=\frac{\langle
Q^2\rangle_c(\rho,\rho)}{t}
\end{equation}
they have shown that
\begin{equation}\label{BDGC}\begin{split}
\mu(z)=\mu\left(2\int^{\rho_0}_{\rho_1}\dd\rho\frac{D(\rho)}{\sigma(\rho)}\frac{1}{z}\right)
\end{split}\end{equation}
To leading order in $\eps$, it is straighforward to see from (\ref{muZRP}) that, for the Zero Range
Process,
\begin{equation}
D(\rho)=1,\;\sigma(\rho)=2\rho
\end{equation}
For the short range process,
\begin{equation}\begin{split}
\text{for pair interaction: }D(\rho)=1+2\eps \rho,\;\sigma(\rho)=2\rho\\
\text{for triplet interaction: }D(\rho)=1+6\eps \rho^2,\;\sigma(\rho)=2\rho\\
\end{split}\end{equation}
thus leading, in both cases, to 
\begin{equation}\begin{split}
\mu(z)=\mu(\frac{\zeta_0}{\zeta_1 z})
\end{split}\end{equation}
Hence formula (\ref{BDGC}) is in perfect agreement with our explicit
computations, not only for zero range processes, but also for the non trivial
short range processes with pair or triplet repulsion.

\section{Final remarks}

We have shown on specific examples that driving a system out of equilibrium
magnifies the differences in the underlying microscopic dynamics. Not only the
density profiles are different, but some dynamical rules lead the
system to a state of local equilibrium, while others let it develop long range
effective interactions. In all cases, however,
our explicit results for the integrated current distribution provide
support for the postulated additivity principle of Bodineau
and Derrida \cite{bodineauderrida}.\\

We should emphasize that the above results have been worked out
by setting up a formalism, based on path integrals, which provides an alternative
both to exact solutions and to the
fluctuating hydrodynamics approach. The path integral formalism
fills a gap in the sense that it allows us to go directly from a
microscopic formulation to macroscopic properties and, furthermore,
it allows formulating approximate approaches in nonequilibrium
settings. Most of our results rely on a
high temperature or virial like expansion, thus providing a intuitive parallel
to the standard techniques of equilibrium statistical mechanics.\\

We believe that several lines could now be explored. First, the present
toolbox allows us to investigate the effect of additional space dimensions at
little extra formal cost (even though calculations will undoubtedly
be more involved). This would help to clearly isolate those feature which are
characteristic of one-dimensional systems from those that generalize to
more realistic ones. An interesting problem arising in higher
dimension is the interplay between
longitudinal and transverse current fluctuations. Second, most of the quantities
studied in the present work are
time-independent, but the present toolbox makes possible
the sudies of time-dependent quantities such as the time-dependent
profile or effective free energy. Third, a natural extension of our formalism
would consist of establishing nonperturbative results in $\eps$. Much in the
same way as in liquid state theory, infinite families of appropriately
chosen Mayer diagrams can be
summed up, and one could investigate which types of Feynman diagrams will
contribute in building up the strongly nonlocal nature of the free energy
functional found in \cite{derridalebowitzspeer1}.
Finally, extending our results for
more complex geometries with more than two particle reservoirs, as suggested in
\cite{bodineauderrida} or done in \cite{blanterbuttiker},
should allow us to identify the universal emerging features and to bring
the theory closer to experiments.\\

\noindent {\bf Ackownledgments.} This research has been partially supported
by the Hungarian Academy
of Sciences (Grant No. OTKA T043734). The authors would like to thank Gunter
Sch\"utz, Henk Hilhorst, Emmanuel Trizac and Alain Barrat for useful
discussions, and further acknowledge Bernard Derrida for numerous
useful suggestions.

\end{document}